\documentclass[journal,
transmag,
]
{IEEEtran}
\usepackage{pifont} 

\usepackage{amssymb}

\usepackage[noadjust]{cite}

\usepackage{graphicx}
\DeclareGraphicsExtensions{.eps}
\graphicspath{{./Images/}} 

\usepackage[squaren, Gray, cdot]{SIunits}

\begin{document}
%
\title{Enhancement of the magnetoelectric effect in multiferroic CoFe$_2$O$_4$/PZT bilayer by induced uniaxial magnetic anisotropy}


\author{\IEEEauthorblockN{Alex Aubert\IEEEauthorrefmark{1},
Vincent Loyau\IEEEauthorrefmark{1},
Fr\'ed\'eric Mazaleyrat\IEEEauthorrefmark{1}, and 
Martino LoBue\IEEEauthorrefmark{1}}
\IEEEauthorblockA{\IEEEauthorrefmark{1}SATIE UMR 8029 CNRS, ENS Paris-Saclay,
Cachan, 94235 FRANCE}\\
DOI : 10.1109/TMAG.2017.2696162
}

%



\IEEEtitleabstractindextext{%
\begin{abstract}
In this study we have compared magnetic, magnetostrictive and piezomagnetic properties of isotropic and anisotropic cobalt ferrite pellets. The isotropic sample was prepared by the ceramic method while the sample exhibiting uniaxial anisotropy was made by reactive sintering using Spark Plasma Sintering (SPS). This technique permits to induce a magnetic anisotropy in cobalt ferrite in the direction of the applied pressure during SPS process. Sample with uniaxial anisotropy revealed a higher longitudinal magnetostriction and piezomagnetism compared to the isotropic sample, but the transversal magnetostriction and piezomagnetism were dramatically reduced. In the case of magnetoelectric layered composite, the magnetoelectric coefficient is directly related to the sum of the longitudinal and transversal piezomagnetic coefficients. These two coefficients being opposite in sign, the use of material exhibiting high longitudinal and low transversal piezomagnetic coefficient (or vice versa) in ME devices is expected to improve the ME effect. Hence, ME bilayer devices were made using isotropic and anisotropic cobalt ferrite stuck with a PZT layer. ME measurements at low frequencies revealed that bilayer with anisotropic cobalt ferrite exhibits a ME coefficient three times higher than a bilayer with isotropic cobalt ferrite. We also investigated the behavior of such composites when excited at resonant frequency.

\end{abstract}

\begin{IEEEkeywords}
Magnetoelectric, Magnetostriction, Magnetic anisotropy, Spark Plasma Sintering, Resonance
\end{IEEEkeywords}}

\maketitle
\date{\today}

\IEEEdisplaynontitleabstractindextext

%
\IEEEpeerreviewmaketitle



\section{Introduction}
\IEEEPARstart{T}{he} magnetoelectric (ME) effect has raised great interest in the recent years because of its potential use in smart electronic application~\cite{scott2012, zhang2016, he2014, abderrahmane2012}. Beside the research for intrinsinc magnetoelectric alloys, relevant advances have been reached in the study of magnetostrictive-piezoelectric heterostructure composite. In this case, the magnetoelectric coupling is due to the magnetic-mechanical-electric transform through the interface between layers. The electromagnetic coupling results from the dynamic mechanical deformation of the ferromagnet which induces a variation of polarization in the piezoelectric layer. Hence, the magnetoelectric effect mainly arises from the dynamic magnetostriction, i.e. the piezomagnetic coefficient $q^m$ of the magnetic material. 

The piezomagnetic coefficient is defined as the slope of the magnetostrictive coefficient $q^m=d\lambda/dH$, and is the meaningful parameter to investigate for sensors and actuators. For magnetoelectric purposes, the magnetoelectric coefficient in the transverse direction $\alpha_{31}$ depends on the sum of the longitudinal $q^m_{11}=d\lambda_{11}/dH$ and the transverse $q^m_{21}=d\lambda_{21}/dH$ piezomagnetic coefficients of the magnetic layer~\cite{loyau2017, srinivasan2003, bichurin2002, filippov2004bi}. This explains why researches on magnetoelectric layered composite are usually focused on good magnetostrictive materials such as Terfenol-D, nickel ferrite or cobalt ferrite associated with lead zirconate titanate (PZT)~\cite{srinivasan2003, wang2005, loyau2017, ryu2002}.

However, magnetic materials used in magnetoelectric devices are usually isotropic. In magnetostrictive properties, this results in a ratio between maximum longitudinal and transverse magnetostriction of 2:1. Moreover, the isotropy of the material implies that longitudinal $\lambda_{11}$ and transverse $\lambda_{21}$ magnetostriction are of opposite sign. The same behavior occurs for piezomagnetic properties, longitudinal $q^m_{11}$ and transverse $q^m_{21}$  piezomagnetic coefficient are opposite in sign and the maximum $|q^m_{11}|$ is two times higher than $|q^m_{21}|$. Thus, by summing up these two coefficient $q^m_{11}+q^m_{21}$, it leads to a piezomagnetic coefficient $q^m_{\sum}$ two times lower than $q^m_{11}$, eventually resulting in a low magnetoelectric coefficient $\alpha_{31}$ since it depends directly on $q^m_{\sum}$. Hence, to increase the magnetoelectric effect, one must enhance $q^m_{\sum}$ which is possible by improving $q^m_{11}$ and keeping $q^m_{21}$ low and vice versa. 

The most common approach to enhance the longitudinal piezomagnetic coefficient ($q^m_{11}$) and decrease the transverse piezomagnetic coefficient ($q^m_{21}$) is to induce uniaxial anisotropy in the material. This can be done in cobalt ferrite by magnetic-annealing~\cite{lo2005, muhammad2012, khaja2012, zheng2011}, which consists in applying a strong magnetic field during annealing between 300 and 400~$^\circ$C. A rearrangement of Co and Fe ions in the crystal structure leads to a uniaxial anisotropy parallel to the direction of the magnetic annealing field. Recently, we proposed~\cite{aubert2017} another technique to induce uniaxial anisotropy in cobalt ferrite, by means of a reaction under uniaxial pressure using Spark Plasma Sintering (SPS). SPS process~\cite{munir2006} is used to make the reaction~\cite{orru2009} and/or the sintering~\cite{cruz2014} of oxide-based materials. During this process, high uniaxial pressure is applied while pulsed electric current heats up the die and the ceramic. It has been shown that using SPS to activate the reaction and the sintering of cobalt ferrite permitted to induce a uniaxial anisotropy along the direction of the applied pressure~\cite{aubert2017}.

In this study, magnetic, magnetostrictive and piezomagnetic properties are compared between cobalt ferrite with uniaxial anisotropy made by SPS, and isotropic cobalt ferrite made by the ceramic method. The ME effect is then compared for CoFe$_2$O$_4$/PZT bilayer using isotropic and anisotropic cobalt ferrite. The advantage of cobalt ferrite with uniaxial anisotropy for magnetoelectric purpose is shown in different frequency ranges.


\section{Experimental Details}
\subsection{Samples fabrication}
Polycristalline cobalt ferrite were prepared by two different methods. In both cases, nanosize oxides ($<$~50~nm) Fe$_2$O$_3$ and Co$_3$O$_4$ (Sigma-Aldrich) were used as precursors in adequate molar ratio. Oxides were mixed in a planetary ball mill during 30~min at 400~rpm, and then grinded during 1~hour at 600~rpm. In the first method, cobalt ferrite was made by the classic ceramic method. The mixture was first calcined at 900~$^\circ$C during 12~hours to form the spinel phase, and then grinded at 550~rpm during 1~hour. After uniaxial compaction at 50~MPa in a cylindrical die of 10~mm diameter, sample was sintered at 1250~$^\circ$C during 10~hours. This sample will be referred as \mbox{CF-CM}. In the second method, Spark Plasma Sintering (SPS) was used to make the reaction and the sintering (reactive sintering) of the cobalt ferrite. The reaction was performed at 500~$^\circ$C for 5~min followed by the sintering stage at 750~$^\circ$C for 3~min, both under a uniaxial pressure of 100~MPa. This sample will be referred as CF-SPS. Both methods resulted in cobalt ferrite with a large majority of spinel phase ($>91~\%$)~\cite{aubert2017}. The final shape of both samples is identical, a disk of 10~mm diameter and 2~mm thick. 

To make magnetoelectric samples, cobalt ferrite disks were bonded on commercial PZT disks (Ferroperm PZ27) of 1~mm thick and 10~mm diameter using silver epoxy (Epotek E4110). The piezoelectric samples are polarized along the thickness direction. The magnetoelectric bilayer is then a disk of thickness 3~mm and 10~mm diameter. 

\subsection{Measurement procedure}
The magnetic measurements were carried out with a vibrating sample magnetometer (VSM, Lakeshore 7400) up to a maximum field of 800~kA/m. The ferrite disks were cut into 8~mm$^3$ cubes to compare the measurements in the three directions of the Cartesian coordinate system (see inset in Figure~\ref{VSM}).
Magnetostriction measurements were performed at room temperature using the strain gauge (Micro-Measurements) method with an electromagnet supplying a maximum field of 700~kA/m. The gauges were bonded on the pellets' surface along the direction (1) and the magnetic field was applied in the directions (1) and (2) in the plane of the disk (see inset in figure~\ref{Magnetostriction}). Hence, longitudinal $\lambda_{11}$ and transverse $\lambda_{21}$ magnetostriction coefficients were obtained.
The magnetoelectric coefficient is measured as function of a continuous magnetic field $H_{DC}$ produced by an electromagnet applied in the transverse direction (1) of the bilayer magnetoelectric sample. A small external AC field is superimposed in the same direction (1~mT, 80~Hz) produced by Helmoltz coils (see inset in Figure~\ref{ME}). The magnetoelectric voltage is measured with a lock-in amplifier (EG\&G Princeton Applied Research Model 5210) having an input impedance of 100~M$\Omega$ for low frequencies. At resonant frequency, the magnetoelectric voltage is measured with an oscilloscope.
Compliances were measured using the ultrasonic velocity measurements along the thickness direction of the disk using the pulse-echo technique (longitudinal and shear waves) at 20~MHz.


\section{Results and discussion}

\subsection{Magnetism}
\begin{figure}[!t]
\centering
\includegraphics[width=0.4\textwidth]{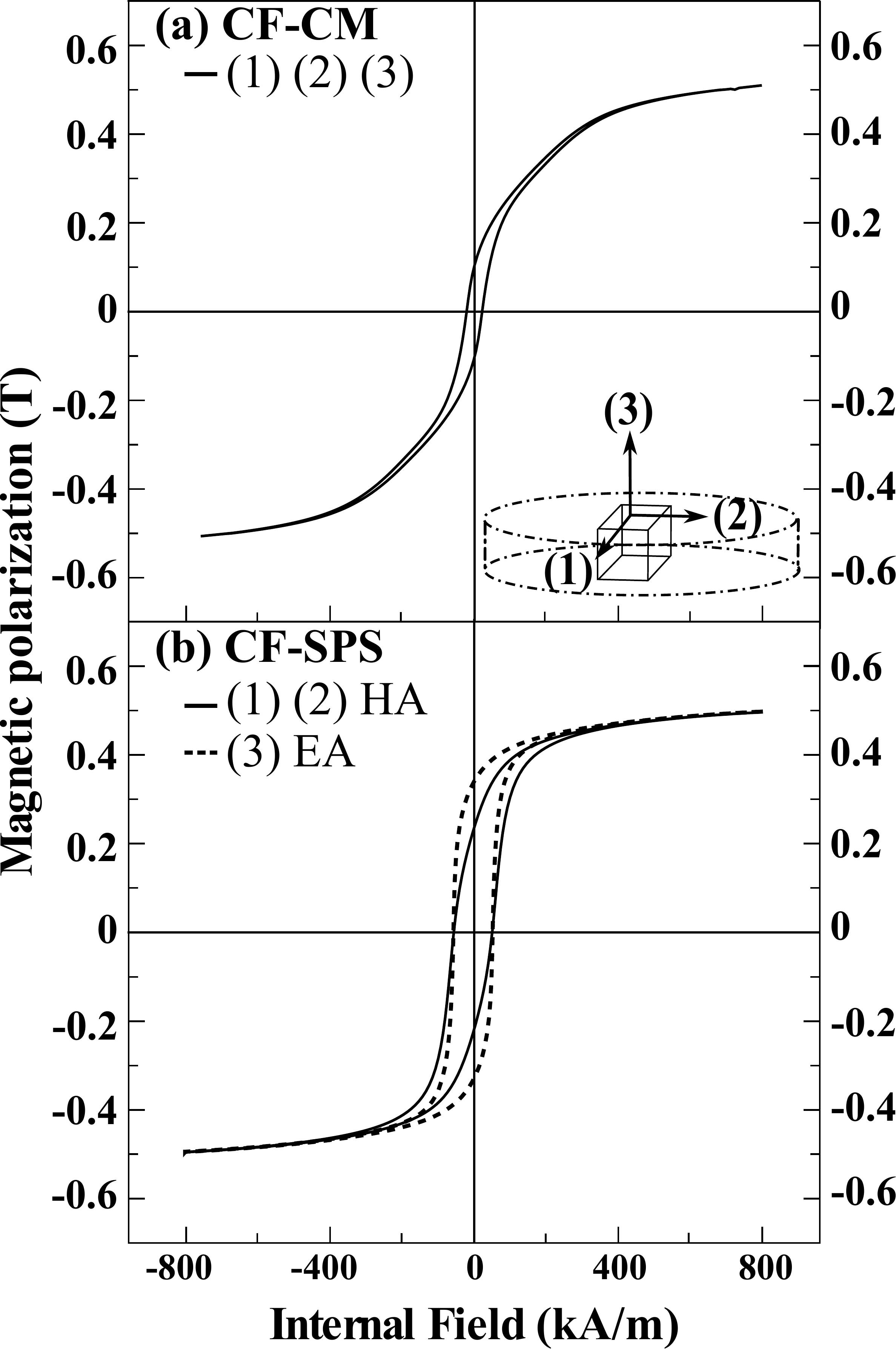}
\caption{Hysteresis loop M-H of samples (a) CF-CM and (b) CF-SPS cut into cube shape. Measurements are done in the three directions of the cube (1), (2) and (3) as represented on the drawing.}
\label{VSM}
\end{figure}

In Figure~\ref{VSM}, we show the magnetic polarization as a function of the internal field, by taking into account the magnetometric demagnetizing factor of a cube ($N_m = 0.2759)$)~\cite{chen2005}. For CF-CM (in Figure~\ref{VSM} (a)), the three hysteresis loops exhibit similar behavior in the three directions of the cube, indicating the isotropy of the material. By opposition, for CF-SPS (in Figure~\ref{VSM} (b)), the measurements show that the remanent magnetization in the direction (3) is higher than for the direction (1) and (2) of the cube. Indeed, the remanent magnetic moment reaches $301$~mT along the easy axis while it is $205$~mT along the hard axis. This behavior indicates a uniaxial anisotropy in the direction (3). This particular direction corresponds to the direction of the pressure applied during the SPS process, confirming reactive sintering under applied pressure as an effective method to induce a uniaxial anisotropy in cobalt ferrite~\cite{aubert2017}.

\subsection{Magnetostriction and Piezomagnetism}
\begin{figure}[!t]
\centering
\includegraphics[width=0.45\textwidth]{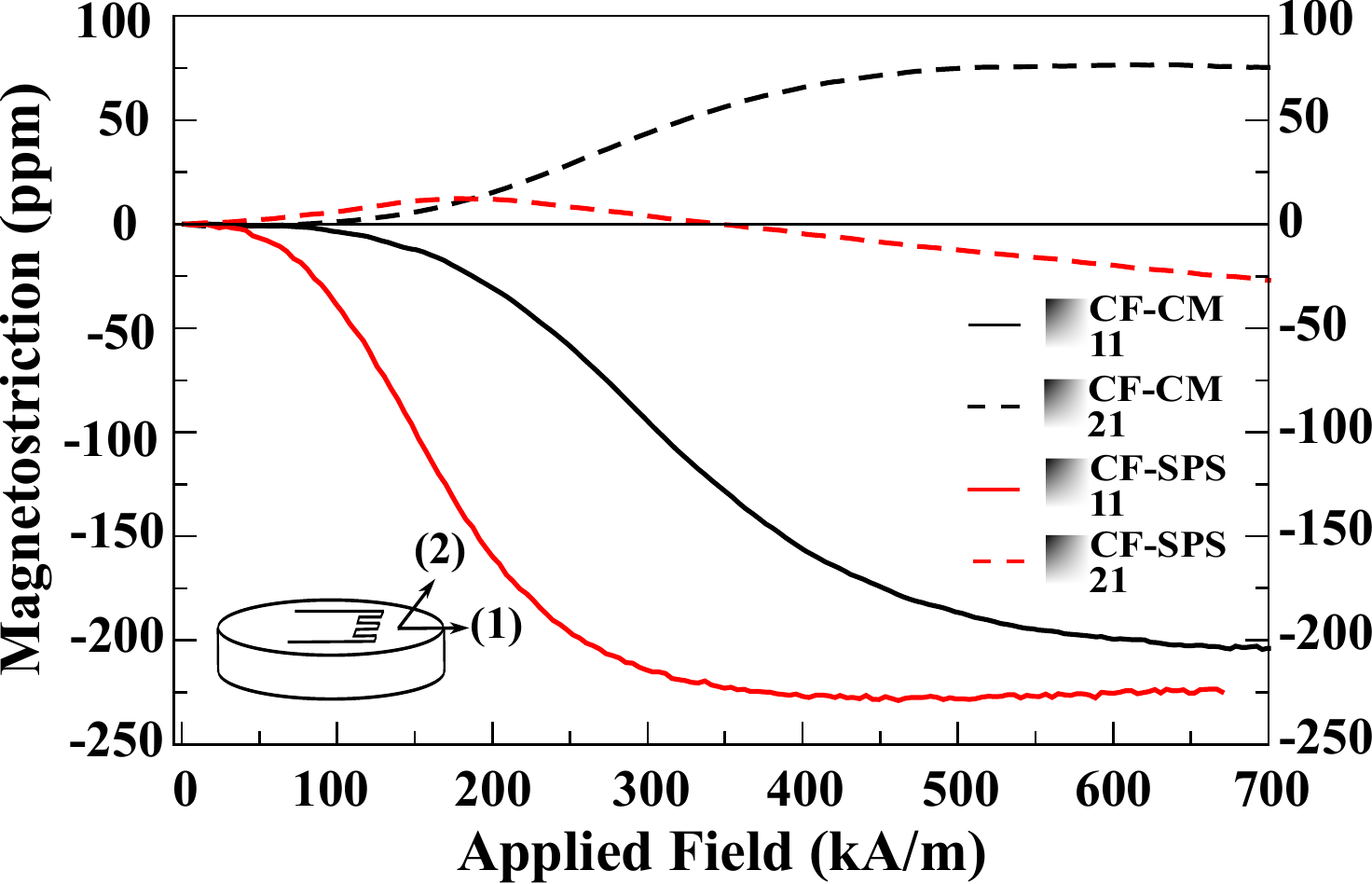}
\caption{Magnetostriction curves of CF-CM and CF-SPS  are represented in black and red respectively. The solid line ($\lambda_{11}$) corresponds to the measurement when the applied field is along the direction (1) and the dash line ($\lambda_{21}$) when the applied field is along the direction (2). The strain gauge is bonded along the direction (1) for all measurements as represented on the drawing.}
\label{Magnetostriction}
\end{figure}

In Figure~\ref{Magnetostriction}, magnetostrictive measurement of CF-CM and CF-SPS in the longitudinal and transverse direction are reported (see inset in Figure~\ref{Magnetostriction}). As expected, cobalt ferrite with uniaxial anisotropy exhibits a different behavior from the isotropic cobalt ferrite. Indeed, CF-CM shows a maximum longitudinal magnetostriction $\lambda_{11}$ of -204 ppm  and a maximum transverse magnetostriction $\lambda_{21}$ of 76 ppm, which are usual values for isotropic CoFe$_2$O$_4$~\cite{zheng2011}. For CF-SPS, the maximum longitudinal magnetostriction has increased to -229 ppm while the transverse magnetostriction has dramatically reduced to 12 ppm and then becomes negative at a given applied field. This type of curves is typical for cobalt ferrite after magnetic annealing showing an induced uniaxial anisotropy~\cite{khaja2012, muhammad2012}. This leads to a ratio between maximum longitudinal and transverse magnetostriction of 19:1 while it is approximatively of 2:1 for isotropic materials. Hence, as expected, the longitudinal magnetostriction of the anisotropic cobalt ferrite is enhanced and the transverse magnetostriction is reduced compared to the isotropic ceramic.

\begin{figure}[!t]
\centering
\includegraphics[width=0.45\textwidth]{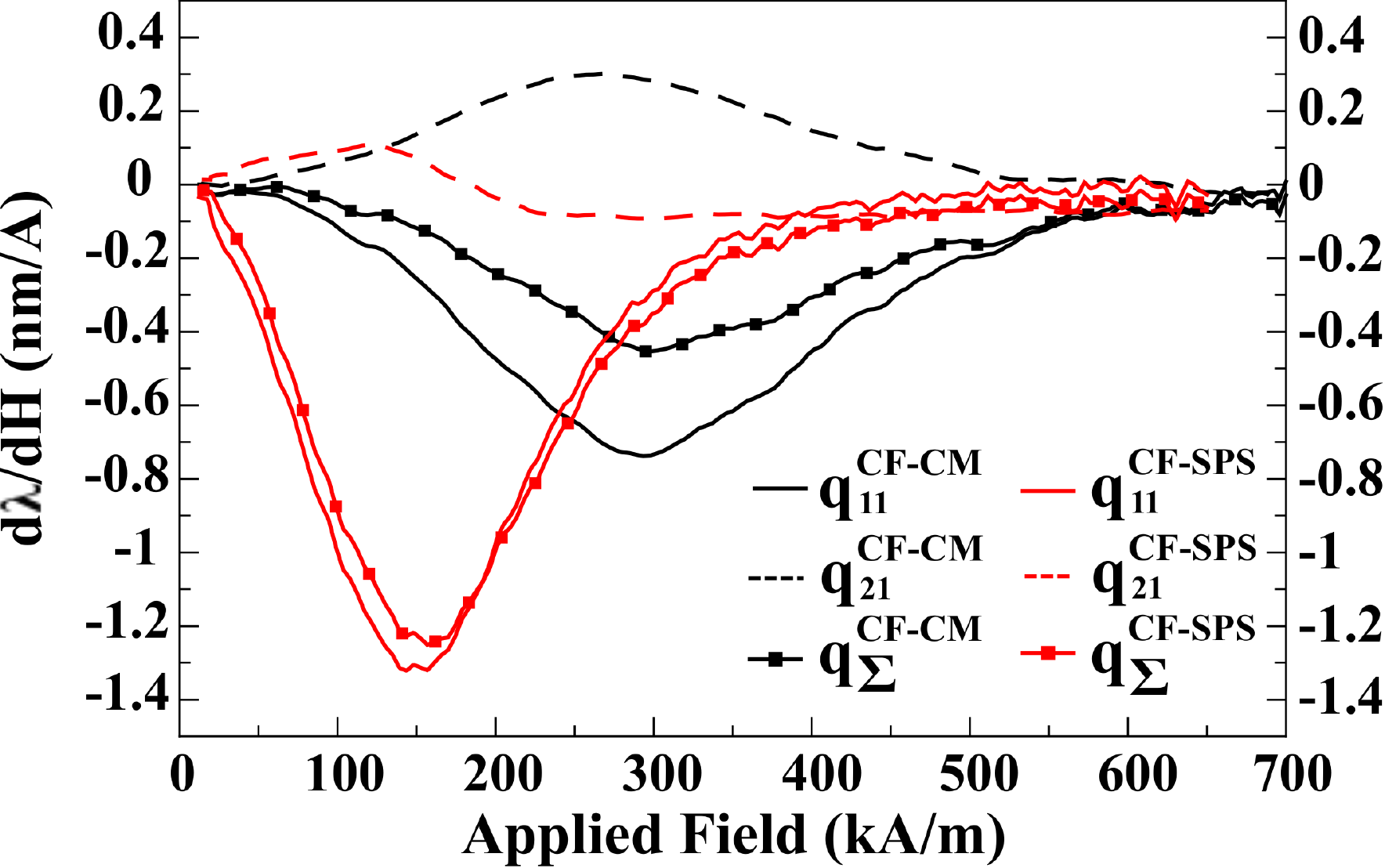}
\caption{Piezomagnetic curves deduced from magnetostrictive measurement for CF-CM and CF-SPS in black and red respectively. The solid line ($q_{11}=d\lambda_{11}/dH$) corresponds to the strain derivative in the direction (1) and the dash line ($q_{21}=d\lambda_{21}/dH$) to the strain derivative in the direction (2). Line with square symbol represents the sum of $q_{11}$ and $q_{21}$.}
\label{Piezomagnetism}
\end{figure}

Introducing uniaxial anisotropy was also found to improve the longitudinal strain derivative $q_{11}^m=d\lambda_{11}/dH$ while reducing the transverse strain derivative  $d_{21}^m$~\cite{muhammad2012, khaja2012}. In Figure~\ref{Piezomagnetism}, the magnetic field derivative of the magnetostrictive curves are represented in the longitudinal and transverse direction for \mbox{CF-CM} and \mbox{CF-SPS}. The sum of both  $q_{\sum}^m = q_{11}^m+q_{21}^m$ is also plotted. The maximum longitudinal strain derivative for \mbox{CF-CM} is -0.73~nm/A while it was increased to -1.3~nm/A for \mbox{CF-SPS}. For the transverse direction, the maximum strain derivative for CF-CM is 0.3~nm/A while it was reduced to 0.1~nm/A for CF-SPS. By summing up these two piezomagnetic coefficient, the strain derivative calculated for CF-CM is -0.45~nm/A while improving to -1.2~nm/A for CF-SPS. As $q_{11}^m$ and $q_{21}^m$ are opposite in sign, the improvement of the sum $q_{\sum}^m$ for cobalt ferrite with induced uniaxial anisotropy \mbox{CF-SPS} is mainly due to the low transverse strain derivative $q_{21}^m$, a direct consequence of the low transverse magnetostriction $\lambda_{21}$ of the sample. Moreover, the applied field required to reach the maximum $q_{\sum}^m$ is reduced for CF-SPS when compared to CF-CM from 300~kA/m to 155~kA/m. Thus, besides increasing $q_{\sum}^m$ by about a factor of three, the uniaxial anisotropy also reduces to half the required applied field to reach the maximum value, which is of great importance to make sensors with high sensitivity while requiring low applied fields. 

\subsection{Magnetoelectric Effect}
\begin{figure}[!t]
\centering
\includegraphics[width=0.45\textwidth]{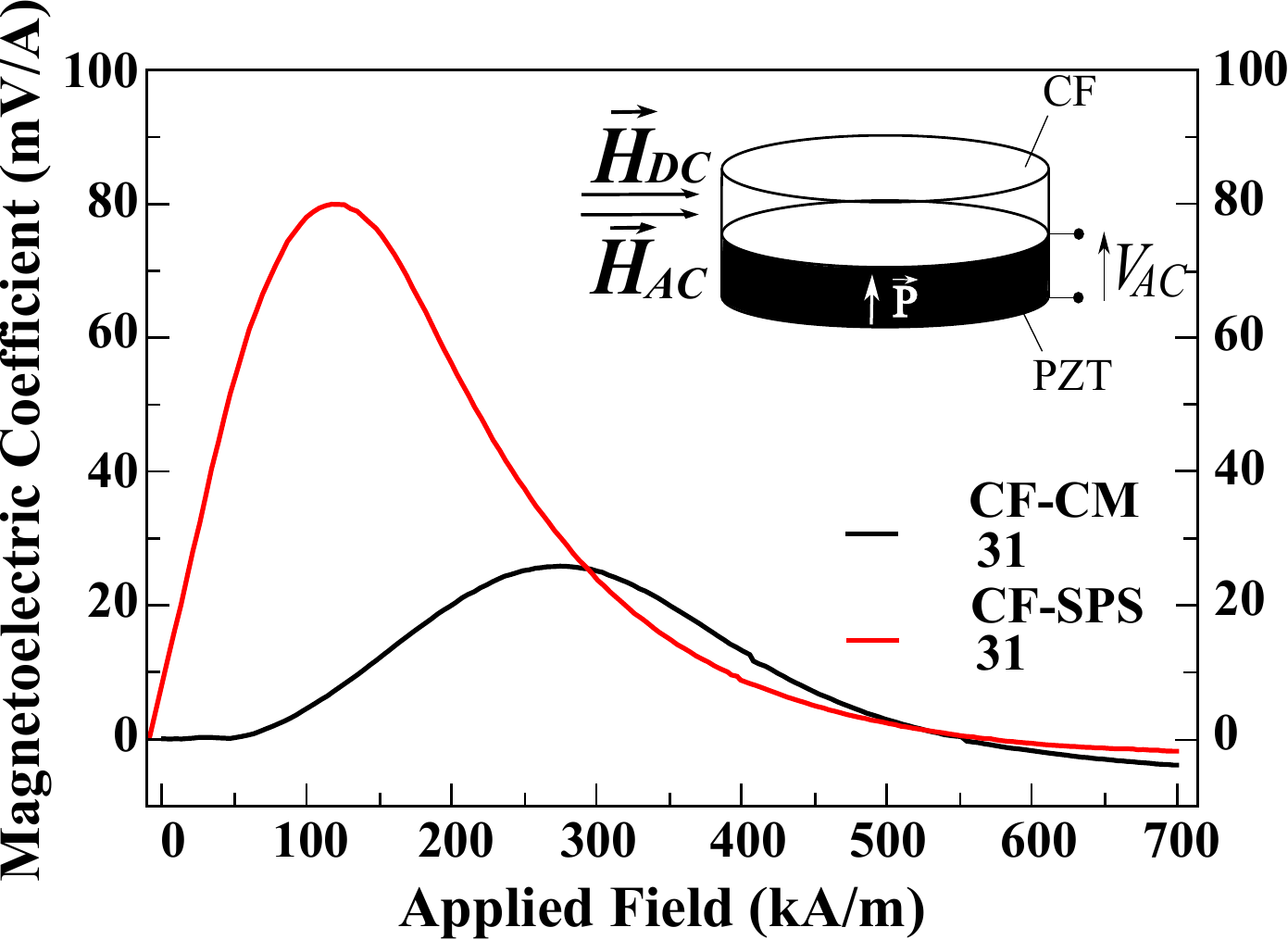}
\caption{Transversal magnetoelectric coefficient $\alpha_{31}$ as function of DC applied field for a bilayer (2/1) of CF-CM/PZT and CF-SPS/PZT in black and red respectively. The AC applied field $H_{AC}$ is of 1 mT at 80 Hz.}
\label{ME}
\end{figure}

To evaluate the potential of these ferrites in magnetoelectric applications, CF-CM and CF-SPS were bonded on PZT disks to obtain magnetoelectric bilayers. Magnetoelectric voltage was measured as function of a DC magnetic field applied in the transverse direction of the bilayer disk while a small AC field (1~mT, 80~Hz) was superimposed in the same direction. Here, low frequency was used to avoid any resonance effect. The transverse magnetoelectric coefficients $\alpha_{31}$ were hence deduced from the piezoelectric voltage measured along the thickness direction. The magnetoelectric setup is represented in the inset of Figure~\ref{ME}. The magnetoelectric coefficient measured for \mbox{CF-CM/PZT} $\alpha_{31}^{CF-CM}$ and \mbox{CF-SPS/PZT} $\alpha_{31}^{CF-SPS}$ are shown in Figure~\ref{ME}. The magnetoelectric effect observed for the bilayer with CF-SPS is about three times higher than the one observed in the bilayer with \mbox{CF-CM}. A maximum magnetoelectric coefficient of 26~mV/A and 80~mV/A are obtained for the CF-CM/PZT and CF-SPS/PZT respectively. Moreover, this maximum value is reached at much lower applied field, 120~kA/m for $\alpha_{31}^{CF-SPS}$ when compared to 275~kA/m for $\alpha_{31}^{CF-CM}$. These results agree well with the piezomagnetic coefficient deduced from the magnetostrictive cruves. Indeed, the magnetoelectric model derived at low frequency~\cite{loyau2017} shows the dependance of $\alpha_{31}$ on $q_{\sum}^m$:
\begin{eqnarray}
\alpha_{31} = \frac{\eta(q^m_{11}+q^m_{21})d^e_{31}}{\epsilon_{33}\big[(s^e_{11}+s^e_{21})+\eta\gamma(s^m_{11}+s^m_{21})\big] -2(d^e_{31})^2}\nonumber
\\
\times \frac{1}{1+N_r\chi}
\label{eq_me}
\end{eqnarray}
where $\eta$ is the mechanical coupling factor, $d^e_{31}$ is the transverse piezoelectric coefficient, $\epsilon_{33}$ is the dielectric permittivity, $s_{ij}$ are the compliance, $\gamma=\frac{\nu_e}{\nu_m}=\frac{t_e}{t_m}$, with $t_e$ and $t_m$ as the thickness of PZT and ferrite respectively, $\chi$ the susceptibility and $N_r$ the demagnetizing factor which depends on the ferrite shape.

Here, both bilayers have the same geometry and mechanical properties. Indeed, compliance were measured for CF-CM, giving : $s_{11}$= 6.74~nm$^2$/N and $s_{21}$= --1.97~nm$^2$/N; and for CF-SPS : $s_{11}$= 6.44~nm$^2$/N and $s_{21}$= -- 1.96~nm$^2$/N. Hence, the meaningful parameter at low frequency behavior should be the piezomagnetic coefficient. This explains why a ratio of three is found between CF-CM and CF-SPS for the maximum magnetoelectric coefficient $\alpha_{31}$, as it was for the piezomagnetic coefficient $q_{\sum}^m$. This also demonstrates that to optimize the transverse magnetoelectric effect $\alpha_{31}$ at low frequency, a low $q_{21}=d\lambda_{21}/dH$ is needed, and a possible way to reach it is to use materials exhibiting uniaxial anisotropy.

\begin{figure}[!t]
\centering
\includegraphics[width=0.45\textwidth]{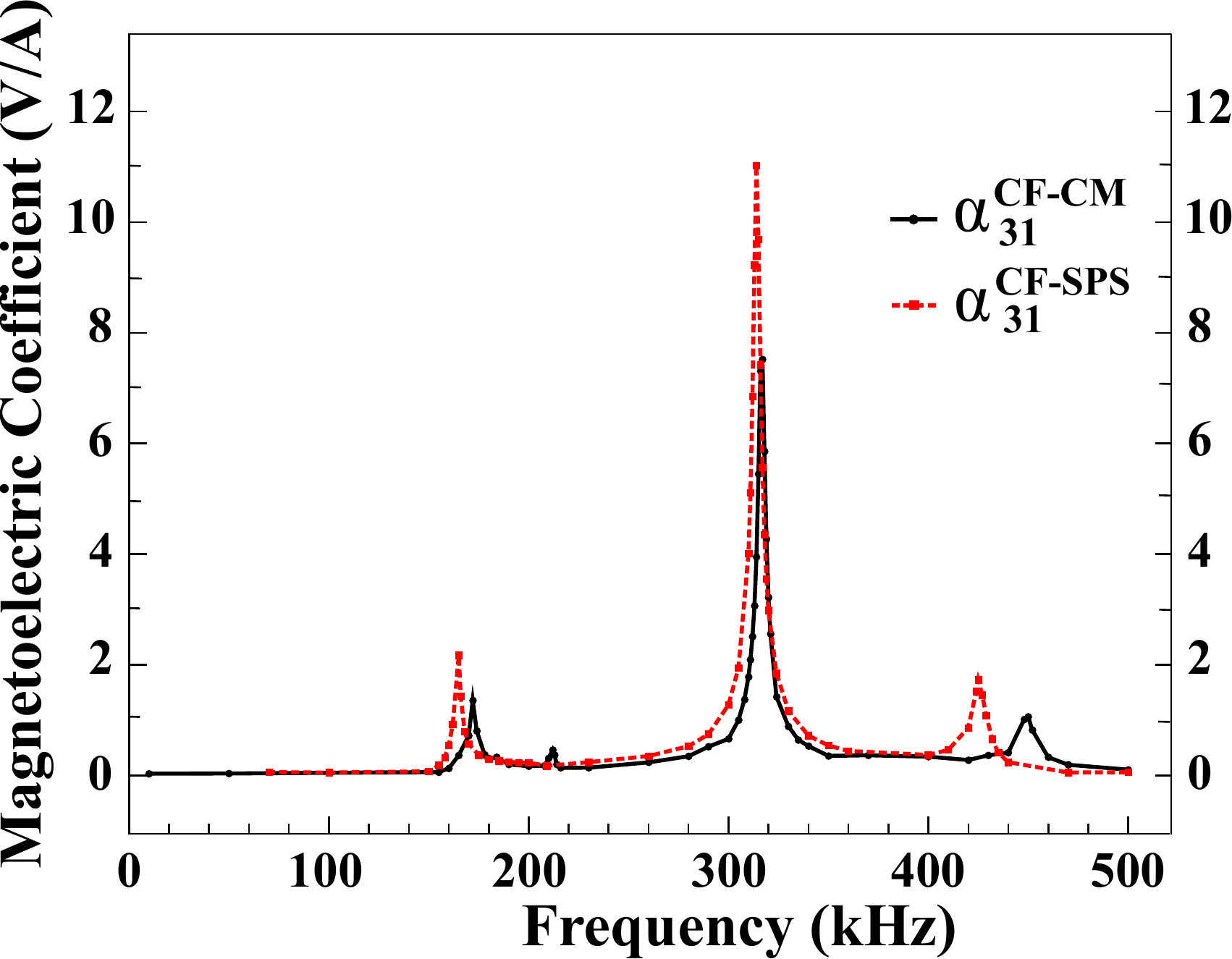}
\caption{Transversal magnetoelectric coefficient $\alpha_{31}$ as function of frequency for a bilayer (2/1) of CF-CM/PZT and CF-SPS/PZT in black and red respectively.}
\label{Res}
\end{figure}

Magnetoelectric measurements were also performed as function of the frequency of the AC magnetic field as plotted in Figure~\ref{Res}. As reported in several papers~\cite{bichurin2003res, filippov2004res, zhang2007}, a bilayer with PZT of 1~cm diameter has an electromechanical resonance (EMR) around 300~kHz. The resonance in ME coefficient occurs when the AC field is tuned to EMR. This is what we observed in the magnetoelectric response of both CF-CM/PZT and CF-SPS/PZT, where the main resonance was found at 317~kHz and 314~kHz respectively (Figure~\ref{Res}). This results in a magnetoelectric coefficient increased to 7.5~V/A for CF-CM/PZT, which is 300 times higher than the coefficient measured at low frequency. For CF-SPS/PZT it was increased to 11~V/A, ``only" 138 times higher when compared to low frequency.


The model developped by Filippov~\cite{filippov2004bi} for a bilayer structure with disks at the resonant frequency highlights the direct dependance of the magnetoelectric coefficient on the sum of $q_{11}^m$ and $q_{21}^m$ as for low frequencies. However, in our case, the ratio between the two bilayers CF-CM/PZT and CF-SPS/PZT for the magnetoelectric coefficient at resonant frequency  is of 1.5 and not 3 as it was at low frequency. At the EMR, mechanical paramaters should be mainly involved in the magnetoelectric coupling compared to the piezomagnetic coefficient. But, as was said before, mechanical properties of CF-CM and CF-SPS are very close, validated by the compliances values. Also, resonant frequency for both bilayer are identical, indicating similar mechanical behavior. So, the meaningful parameter at EMR in this case seems to be either the damping factor~\cite{filippov2004bi}, also named mechanical loss factor~\cite{bichurin2003res}, or the mechanical coupling coefficient. These parameters might depend on the microstructure of the cobalt ferrite. Here, CF-CM has lower relative density (90~\%) than CF-SPS (97~\%) because SPS sintering allows very dense materials~\cite{aubert2017}. Moreover, CF-CM has much larger grain size ($\sim$~\unit{4}{\micro\meter}) than CF-SPS ($<$~\unit{100}{\nano\meter}), because SPS permits very short time process, hence the grain growth does not occur~\cite{aubert2017, orru2009}. These microstructure properties could affect the damping factor, and/or the mechanical coupling coefficient, explaining the difference in amplitude found for the magnetoelectric coefficient between the two bilayers CF-CM/PZT and CF-SPS/PZT at the resonant frequency.

Some minor peaks are also present at other frequencies such as 172~kHz, 212~kHz and 448~kHz for CF-CM/PZT and 165~kHz and 425~kHz for CF-SPS/PZT. These peaks might be a consequence of the structure used here, which is a bilayer. In fact, if the mechanical coupling at the interface is not perfect, it can results in a minor improvement of the magnetoelectric effect at other frequencies than EMR for bilayers~\cite{filippov2007}.

\section{Conclusion}
In summary, magnetic, magnetostrictive and piezomagnetic properties are compared for isotropic and anisotropic cobalt ferrite disks. Isotropic behavior was observed for cobalt ferrite made by the ceramic method while anisotropic properties were found for cobalt ferrite made by reactive sintering at Spark Plasma Sintering. This has a direct effect on the magnetostrictive behavior and particularly in the piezomagnetic coefficient, were the maximum $q_{\sum}^m=d\lambda_{\sum}/dH$ obtained was three times higher for \mbox{CF-SPS} than for CF-CM and for a lower magnetic applied field. As the magnetoelectric effect is expected to depend mainly on the sum $q_{\sum}^m$, the maximum magnetoelectric coefficient obtained at low frequency for the bilayer \mbox{CF-SPS/PZT} is three times higher than for \mbox{CF-CM/PZT}. This result points out the importance of investigating at both piezomagnetic coefficient $q_{11}^m$ and $q_{21}^m$ to determine if a magnetic material has good magnetoelectric potential. Measurement at the resonant frequency show that magnetoelectric effect for anisotropic coblat ferrite was 1.8 times higher than for isotropic cobalt ferrite. Thus, this study validate the recent interest in making cobalt ferrite with induced uniaxial anisotropy for magnetoelectric purpose in all frequency range.


\bibliographystyle{IEEEtran}

\begin{thebibliography}{10}
\providecommand{\url}[1]{#1}
\csname url@samestyle\endcsname
\providecommand{\newblock}{\relax}
\providecommand{\bibinfo}[2]{#2}
\providecommand{\BIBentrySTDinterwordspacing}{\spaceskip=0pt\relax}
\providecommand{\BIBentryALTinterwordstretchfactor}{4}
\providecommand{\BIBentryALTinterwordspacing}{\spaceskip=\fontdimen2\font plus
\BIBentryALTinterwordstretchfactor\fontdimen3\font minus
  \fontdimen4\font\relax}
\providecommand{\BIBforeignlanguage}[2]{{%
\expandafter\ifx\csname l@#1\endcsname\relax
\typeout{** WARNING: IEEEtran.bst: No hyphenation pattern has been}%
\typeout{** loaded for the language `#1'. Using the pattern for}%
\typeout{** the default language instead.}%
\else
\language=\csname l@#1\endcsname
\fi
#2}}
\providecommand{\BIBdecl}{\relax}
\BIBdecl

\bibitem{scott2012}
J.~F. Scott, ``Applications of magnetoelectrics,'' \emph{J. Mater. Chem.},
  vol.~22, pp. 4567--4574, 2012.

\bibitem{zhang2016}
S.~Zhang, M.~Zhang, and S.~W. Or, ``Bidirectional current-voltage converter
  based on coil-wound, intermagnetically biased, heterostructured
  magnetoelectric ring,'' \emph{IEEE Trans. Magn.}, vol.~52, no.~7, pp. 1--5,
  2016.

\bibitem{he2014}
W.~He, P.~Li, Y.~Wen, J.~Zhang, A.~Yang, and C.~Lu, ``Energy harvesting from
  two-wire power cords using magnetoelectric transduction,'' \emph{IEEE Trans.
  Magn.}, vol.~50, no.~8, pp. 1--5, 2014.

\bibitem{abderrahmane2012}
A.~Abderrahmane, S.~Koide, S.~I. Sato, T.~Ohshima, A.~Sandhu, and H.~Okada,
  ``Robust hall effect magnetic field sensors for operation at high
  temperatures and in harsh radiation environments,'' \emph{IEEE Trans. Magn.},
  vol.~48, pp. 4421--4423, 2012.

\bibitem{loyau2017}
V.~Loyau, A.~Aubert, M.~LoBue, and F.~Mazaleyrat, ``Analytical modeling of
  demagnetizing effect in magnetoelectric ferrite/pzt/ferrite trilayers taking
  into account a mechanical coupling,'' \emph{Journal of Magnetism and Magnetic
  Materials}, vol. 426, pp. 530 -- 539, 2017.

\bibitem{srinivasan2003}
G.~Srinivasan, E.~T. Rasmussen, and R.~Hayes, ``Magnetoelectric effects in
  ferrite-lead zirconate titanate layered composites: The influence of zinc
  substitution in ferrites,'' \emph{Phys. Rev. B}, vol.~67, p. 014418, 2003.

\bibitem{bichurin2002}
M.~I. Bichurin, V.~M. Petrov, and G.~Srinivasan, ``Theory of low-frequency
  magnetoelectric effects in ferromagnetic-ferroelectric layered composites,''
  \emph{J. Appl. Phys.}, vol.~92, pp. 7681--7683, 2002.

\bibitem{filippov2004bi}
D.~A. Filippov, ``Theory of magnetoelectric effect in
  ferromagnetic-piezoelectric bilayer structures,'' \emph{Tech. Phys. Lett.},
  vol.~30, pp. 983--986, 2004.

\bibitem{wang2005}
Y.~Wang, H.~Yu, M.~Zeng, J.~Wan, M.~Zhang, J.-M. Liu, and C.~Nan, ``Numerical
  modeling of the magnetoelectric effect in magnetostrictive piezoelectric
  bilayers,'' \emph{Appl. Phys. A}, vol.~81, no.~6, pp. 1197--1202, 2005.

\bibitem{ryu2002}
J.~Ryu, S.~Priya, K.~Uchino, and H.-E. Kim, ``Magnetoelectric effect in
  composites of magnetostrictive and piezoelectric materials,'' \emph{J.
  Electroceram.}, vol.~8, no.~2, pp. 107--119, 2002.

\bibitem{lo2005}
C.~C.~H. Lo, A.~P. Ring, J.~E. Snyder, and D.~C. Jiles, ``Improvement of
  magnetomechanical properties of cobalt ferrite by magnetic annealing,''
  \emph{IEEE Trans. Magn.}, vol.~41, no.~10, pp. 3676--3678, 2005.

\bibitem{muhammad2012}
A.~Muhammad, R.~Sato-Turtelli, M.~Kriegisch, R.~Grössinger, F.~Kubel, and
  T.~Konegger, ``Large enhancement of magnetostriction due to compaction
  hydrostatic pressure and magnetic annealing in cofe2o4,'' \emph{J. Appl.
  Phys.}, vol. 111, no.~1, p. 013918, 2012.

\bibitem{khaja2012}
K.~Khaja~Mohaideen and P.~A. Joy, ``High magnetostriction and coupling
  coefficient for sintered cobalt ferrite derived from superparamagnetic
  nanoparticles,'' \emph{Appl. Phys. Lett.}, vol. 101, no.~7, p. 072405, 2012.

\bibitem{zheng2011}
Y.~X. Zheng, Q.~Q. Cao, C.~L. Zhang, H.~C. Xuan, L.~Y. Wang, D.~H. Wang, and
  Y.~W. Du, ``Study of uniaxial magnetism and enhanced magnetostriction in
  magnetic-annealed polycrystalline cofe2o4,'' \emph{J. Appl. Phys.}, vol. 110,
  no.~4, p. 043908, 2011.

\bibitem{aubert2017}
A.~Aubert, V.~Loyau, F.~Mazaleyrat, and M.~LoBue, ``Uniaxial anisotropy and
  enhanced magnetostriction of cofe2o4 induced by reaction under uniaxial
  pressure with sps,'' \emph{J. Eur. Ceram. Soc.}, 2017
  http://doi.org/10.1016/j.jeurceramsoc.2017.03.036.

\bibitem{munir2006}
Z.~A. Munir, U.~Anselmi-Tamburini, and M.~Ohyanagi, ``The effect of electric
  field and pressure on the synthesis and consolidation of materials: A review
  of the spark plasma sintering method,'' \emph{J. Mater. Sci.}, vol.~41,
  no.~3, pp. 763--777, 2006.

\bibitem{orru2009}
R.~Orr\'u, R.~Licheri, A.~M. Locci, A.~Cincotti, and G.~Cao,
  ``Consolidation/synthesis of materials by electric current activated/assisted
  sintering,'' \emph{Mater. Sci. Eng. R-Reports}, vol.~63, no. 4–6, pp.
  127--287, 2009.

\bibitem{cruz2014}
B.~Cruz-Franco, T.~Gaudisson, S.~Ammar, A.~Bolarin-Miro, F.~Sanchez~de Jesus,
  F.~Mazaleyrat, S.~Nowak, G.~Vazquez-Victorio, R.~Ortega-Zempoalteca, and
  R.~Valenzuela, ``Magnetic properties of nanostructured spinel ferrites,''
  \emph{IEEE Trans. Magn.}, vol.~50, pp. 1--6, 2014.

\bibitem{chen2005}
D.-X. Chen, E.~Pardo, and A.~Sanchez, ``Demagnetizing factors for rectangular
  prisms,'' \emph{IEEE Trans. Magn.}, vol.~41, pp. 2077--2088, 2005.

\bibitem{bichurin2003res}
M.~I. Bichurin, D.~A. Filippov, V.~M. Petrov, V.~M. Laletsin, N.~Paddubnaya,
  and G.~Srinivasan, ``Resonance magnetoelectric effects in layered
  magnetostrictive-piezoelectric composites,'' \emph{Phys. Rev. B}, vol.~68, p.
  132408, 2003.

\bibitem{filippov2004res}
D.~A. Filippov, M.~I. Bichurin, V.~M. Petrov, V.~M. Laletin, N.~N. Poddubnaya,
  and G.~Srinivasan, ``Giant magnetoelectric effect in composite materials in
  the region of electromechanical resonance,'' \emph{Tech. Phys. Lett.},
  vol.~30, no.~1, pp. 6--8, 2004.

\bibitem{zhang2007}
N.~Zhang, D.~Liang, T.~Schneider, and G.~Srinivasan, ``Is the magnetoelectric
  coupling in stickup bilayers linear?'' \emph{J. Appl. Phys.}, vol. 101,
  no.~8, p. 083902, 2007.

\bibitem{filippov2007}
D.~A. Filippov, U.~Laletsin, and G.~Srinivasan, ``Resonance magnetoelectric
  effects in magnetostrictive-piezoelectric three-layer structures,'' \emph{J.
  Appl. Phys.}, vol. 102, no.~9, p. 093901, 2007.

\end{thebibliography}

\end{document}